\documentclass[multphys,vecphys]{svmult}

\usepackage{makeidx}         
\usepackage{graphicx}        
\usepackage{multicol}        
\usepackage[bottom]{footmisc}

\usepackage{amssymb}
\usepackage{amsmath}

\makeindex             

\begin{document}

\title*{Size and Orientation of the `Z' in ZRGs}
\author{Christian Zier}
\institute{Raman Research Institute, Bangalore (India)
\texttt{chzier@rri.res.in}}
%
%
\maketitle

\section{Introduction}
\textsf{X}-shaped radio galaxies (XRGs) form a small subclass of radio
galaxies with two misaligned pairs of radio lobes and radio luminosities close
to the FR~I/II transition. Assuming that the jets are aligned with the spin of
the black hole (BH) at their origin these sources can be explained by rapid
realignment of the jet: The coalescence of a supermassive black hole binary
(BHB), a product of a galaxy merger, quickly realigns the jet ($\lesssim
10^7\,\mathrm{yr}$), with the spin of the merged black hole being dominated by
the orbital angular momentum of the BHB \cite{cz:rottmann01, cz:zier01,
cz:zier02, cz:chirvasa02}. In at least two XRGs (NGC~326 and 3C~52) the
ridges of the secondary lobes have been observed to be offset from each other
laterally by about their width, hence showing a \textsf{Z}-shaped symmetry
about the nucleus. As the galaxy merger proceeds and the captured galaxy
spirals to the common center it induces a rotational stream-field in the ISM
on large scales. If its trajectory passes through the polar regions of the
primary galaxy, the ram pressure of the rotating ISM bends the original jet
into \textsf{Z}-shape \cite{cz:gopal03}.

\section{Deprojection of ZRGs \& Results}
Because the pre-merger spin of the primary BH and the orbital angular momentum
of the merging BHB are not correlated the distribution of the angle between
them is $\propto\sin\theta_\mathrm{jet}$, peaking at $90^\circ$. Since the
post-merger spin is dominated by the orbital angular momentum, the angle
between pre- and post-merger jet follows the same distribution. This
corresponds to the requirement that in \textsf{Z}-shaped radio galaxies (ZRGs)
the secondary's trajectory passes through the polar regions of the primary
galaxy. Spiralling inwards the ram pressure of the rotating ISM balances that
of the jet in a distance $r$ and the jet, with a power close to the FR~I/II
transition, will be bent into a \textsf{Z}-shape.

With the observed projected lateral offset of the ridges $y_\mathrm{p}$ in the
known distance $D$ of the galaxy we can determine the angle
$\theta_\mathrm{jet}$ between the pre- and post-merger jets in dependence on
the bending radius $r$ and the angle $\theta_\mathrm{los}$ between our line of
sight (LOS) and the post-merger jet. There are three possible orientations of
the jets and the LOS relative to each other. The requirement of
$\theta_\mathrm{jet}$ and $\theta_\mathrm{los}$ to be large ($\gtrsim
60^\circ$) imposes certain limits on the bending radius: Only for $r\gtrsim
y_\mathrm{p}/2$ first the orientation with both, the LOS and the receding
part of the pre-merger jet in the same hemisphere, which is defind by the
post-merger jet as polar axis, becomes possible. At $r\approx y_\mathrm{p}
\sqrt{3}/2$ also both other orientations enter the region of allowed angles,
but are less likely. For NGC~326 and 3C~52 we obtain the limiting radii
$10,\,17\,\mathrm{kpc}$ and $25,\,43\,\mathrm{kpc}$ respectively
\cite{cz:zier05}. This result, based on geometrical arguments only, is
consistent with the location of $\sim 50 \, \mathrm{kpc}$ where the ram
pressure of a FR~I/II-jet equates that of the rotating ISM. The peoperties of
the gas stream which are required to bend the jet in \textsf{Z}-shape
($M\approx 10^9\,M_\odot, v\approx 200\,\mathrm{km/s}, r\approx 30 \text{--}
100\,\mathrm{kpc}$) are also in very good agreement with observations and
numerical simulations of galaxy mergers (see references in
\cite{cz:zier05}).

\section{Discussion \& Conclusions}
Our results strongly support to the merger model for XRGs, and ZRGs as a
special subclass with $\theta_\mathrm{jet}\approx90^\circ$. They also
strengthen the conjecture that a jet is aligned with the spin of the BH at its
base and that the jet flips into the direction of the orbital angular momentum
of the pre-merger BHB. We also could restrict the bending radius to the range
$30\text{--}100\,\mathrm{kpc}$. Another important result is that most likely
the LOS and the receding part of the pre-merger jet are in the same hemisphere
whose polar axis is defined by the post-merger jet.  Knowing the correct
orientation we also know the sense of rotation and consequently the direction
of the spin of the post-merger BH. These conclusions have also an important
impact on the conjecture favoured by some authors that after a merger of two
galaxies the decay of the BHB stalls due to the depletion of the so-called
loss cone. In contradiction the existence of XRGs and ZRGs proves that the
binary has merged. In ZRGs it probably merges on timescales of some
$10^8\,\mathrm{yr}$ after the bending of the jet in a distance of about
$50\,\mathrm{kpc}$. Thus, in a way, the bending starts a stop watch for the
rest of the merger. For figures and detailed calculations see
\cite{cz:zier05}.

%
%
%

%
%

%
%



\printindex
\end{document}